# Incorporating Cyclic Group Equivariance into Deep Learning for Reliable Reconstruction of Rotationally Symmetric Tomography Systems

Yaogong Zhang, Fang-Fang Yin, Lei Zhang*

Duke Kunshan University, Jiangsu, China

**Abstract**— Rotational symmetry is a defining feature of many tomography systems, including computed tomography (CT) and emission computed tomography (ECT), where detectors are arranged in a circular or periodically rotating configuration. This study revisits the image reconstruction process from the perspective of hardware-induced rotational symmetry and introduces a cyclic group equivariance framework for deep learning-based reconstruction. Specifically, we derive a mathematical correspondence that couples cyclic rotations in the projection domain to discrete rotations in the image domain, both arising from the same cyclic group inherent in the hardware design. This insight also reveals the uniformly distributed circular structure of the projection space. Building on this principle, we provide a cyclic rotation equivariant convolution design method to preserve projection domain symmetry and a cyclic group equivariance regularization approach that enforces consistent rotational transformations across the entire network. We further integrate these modules into a domain transform reconstruction framework and validate them using digital brain phantoms, training on discrete models and testing on more complex and realistic "fuzzy" variants. Results indicate markedly improved generalization and stability, with fewer artifacts and better detail preservation, especially under data distribution deviation. These findings highlight the potential of cyclic group equivariance as a unifying principle for tomographic reconstruction in rotationally symmetric systems, offering a flexible and interpretable solution for scenarios with limited data.

## Introduction

Medical image reconstruction plays a central role in modern healthcare, enabling the visualization of internal anatomical structures and functions from raw sensor data. High-quality reconstruction is paramount for accurate disease diagnosis, treatment planning, and therapy evaluation. In recent years, deep learning has introduced transformative opportunities for improving image reconstruction, addressing challenges such as incomplete sampling, noise suppression, and accelerated imaging [1]. Among various deep learning paradigms, domain transform reconstruction, which replaces the entire reconstruction pipeline with a single data-driven model, has attracted particular attention [2], [3]. This approach proves especially pertinent in rotationally symmetric tomography systems such as CT and ECT [4]–[11], where the reconstruction problem is inherently ill-posed. By eliminating traditional intermediate steps, domain-transform methods can model the complex, nonlinear relationship between projection and image domains, thereby offering an appealing end-to-end framework.

However, deep learning reconstruction methods still face persistent stability and generalizability challenges, stemming from limited datasets and variability in patient anatomy [1], [12], [13]. Although generating synthetic datasets with digital phantoms is a common strategy [5], [7], discrepancies between simulated and real-world data often impede clinical translation. These challenges underscore the necessity for deep learning-based reconstruction to evolve toward more reliable and interpretable models, capable of achieving consistent and stable performance.

In deep learning, equivariant neural networks offer a promising avenue by exploiting symmetries to capture the inherent structure of data, improving generalizability and data efficiency—particularly in scenarios with scarce training samples [14]–[18]. For instance, convolutional neural networks (CNNs) naturally implement translational equivariance, a key factor behind robust feature extraction in classification or segmentation [19]. More advanced architectures exploit rotational or spherical symmetries, providing further gains in data efficiency and model interpretability within medical imaging applications. For example, Dense Steerable Filter CNNs exploit rotational symmetry in histology images, reducing the number of parameters and mitigating overfitting [20]. 3D Group CNNs have demonstrated superior data efficiency for pulmonary nodule detection, achieving higher accuracy and significantly reducing false positives, even with small datasets[21]. Similarly, Spherical CNNs excel in image denoising tasks, minimizing dataset requirements while preserving image quality [22]. Roto-translation equivariant networks, leveraging SE(2)-group convolution layers, enhance mitosis detection, nuclei segmentation, and tumor detection in histopathology images by encoding rotational and translational symmetry, ensuring robust performance across orientations [23]. Furthermore, SE(3)-equivariant networks have advanced motion tracking in brain MRI by accurately modeling rigid-body transformations and incorporating denoising capabilities, particularly in fetal MRI [24]. Collectively, these examples illustrate how equivariance can address critical challenges in medical imaging, enhancing performance and reliability.

Despite these progresses, equivariant neural networks have been applied far less frequently to image reconstruction. In most conventional image processing tasks, both inputs and outputs reside in the same space (e.g., pixel-aligned images), allowing straightforward equivariance definitions. Similarly, existing hybrid reconstruction methods primarily use roto-translational equivariance to regularize reconstructed images [25], [26], with a focus confined to the image domain. However, unlike common image processing tasks, image reconstruction involves a more complex pipeline that spans multiple domains [1]. This cross-domain nature introduces additional challenges in defining and applying equivariance relationships throughout the reconstruction process.

The image reconstruction problem is the inverse problem of the forward imaging process, constrained by the physical properties of the imaging system. In many tomographic systems—such as CT, ECT, and optoacoustic tomography (OAT)—rotational symmetry emerges naturally from the circular arrangement of detectors or periodic rotational data acquisition. This symmetry not only dictates the forward imaging process but also shapes the nature of transformations involved in image reconstruction, providing the foundational motivation for this study.

Leveraging the systems' rotational symmetry, we derive a mathematical correspondence that spans the entire reconstruction pipeline, linking cyclic rotation transformations in the projection domain to discrete rotation transformations in the image domain. Both arise from the same cyclic group defined by the hardware's symmetry, leading to what we term *cyclic group equivariance* (CGE). Notably, the equivariance also reveals a circular structure inherent to the projection data, reflecting the uniform angular distribution imposed by the detectors.

Building upon this principle, we propose two key methods to integrate it into the domain transform reconstruction process. First, we propose a design method for the cyclic rotation equivariant convolution (CREC), which strictly enforces symmetry within the projection domain, and demonstrates one simple implementation within the framework. Second, we introduce a cyclic group equivariance regularization (CGER) strategy to maintain transformation consistency throughout the learning process. Experimental results demonstrate that embedding these methods into domain-transform reconstruction frameworks effectively leverages the unique properties of cyclic group equivariance, mitigating artifacts and structural distortions, especially when the real data distribution deviates from that of the training data. This work thus positions cyclic group equivariance as a unifying principle for rotationally symmetric systems, offering a theoretical and practical foundation for advancing deep learning models in medical image reconstruction.

## Methodology

This section provides both the theoretical underpinnings and the practical framework for incorporating cyclic group equivariance into deep learning-based reconstruction. We begin with a concise review of group theory, which underlies the concept of equivariance. We then connect rotational symmetry in tomographic hardware to a discrete cyclic group structure that simultaneously governs the projection and image domains. Building on this equivariance, we introduce a cyclic rotation equivariant convolution (CREC) to preserve symmetry in the projection space, and a cyclic group equivariance regularization (CGER) to enforce consistent transformations across the entire network.

### Mathematical Background: Groups and Group Representations

A *group* $G$ is defined as a set of elements equipped with a binary operation that satisfies four fundamental properties: closure, associativity, the existence of an identity element $e$, and the existence of an inverse for each element. To illustrate these properties more concretely, we focus on a specific type of group: the cyclic group. A *cyclic group* is a mathematical group in which every element can be expressed as a power (or repeated application) of a single element, known as the generator. For a cyclic group of order $N$, the elements are:

$$C_N = \{e, g, g^2, \dots, g^{N-1}\} = \langle g \rangle, \tag{1}$$

Where $g^N = e$. The inverse of $g^k$ is $g^{N-k}$, ensuring the existence of inverses. Importantly, the closure property ensures that the operation of any two elements in $C_N$ remains in it:

$$g^m * g^n = g^{(m+n) \bmod N}, \forall m, n \in \mathbb{Z}_N. \tag{2}$$

A familiar example of a cyclic group is the integers modulo $N$, denoted $\mathbb{Z}_N$, with addition modulo $N$ as the operation. For instance, the 12-hour clock system corresponds to $\mathbb{Z}_{12}$, where the elements are $\{0,1,2,\dots,11\}$. The group operation is addition modulo $12$, the identity element is $0$, and the inverse of $k$ is $(12 - k) \bmod 12$.

Another fundamental concept in group theory is that of a *subgroup*. A subgroup $H$ of a group $G$ is a subset of $G$ that satisfies the group axioms under the same binary operation as $G$. For cyclic groups, an important property is: All subgroups of a cyclic group $C_N$ are cyclic. A subgroup $C_M$ exists if and only if $M \mid N$ (i.e., $M$ divides $N$), and it has order $M$. For example, the cyclic group $\mathbb{Z}_{12}$ has subgroups of orders $M = 1,2,3,4,6,12$. The subgroup of order $M = 4$, denoted $C_4$, consists of elements $\{0,3,6,9\}$.

*Group representations* provide a concrete realization of abstract group elements by expressing them as linear transformations or matrices acting on vector spaces, enabling their application in practical computations and analyses. For cyclic group $C_N$, one common representation is in the form of rotation matrices:

$$R_n = \begin{bmatrix} \cos\frac{2\pi n}{N} & -\sin\frac{2\pi n}{N} \\ \sin\frac{2\pi n}{N} & \cos\frac{2\pi n}{N} \end{bmatrix}, n \in \mathbb{Z}_N. \tag{3}$$

Here, $R_n$ represents a counterclockwise rotation by $n \cdot \frac{2\pi}{N}$ radians. For example, in the same case of a 12-hour clock system, the discrete rotation of the hour hand can be interpreted as a group representation of $\mathbb{Z}_{12}$, where each element of the cyclic group is associated with a discrete $\frac{2\pi}{12}$ rotation.

Building on the concept of group representations, we can now provide a rigorous definition of equivariance. Let $G$ be a group, and let $T$ and $T'$ be two representations of $G$ acting on spaces $X$ and $Y$, respectively. A mapping $f: X \to Y$ is said to be equivariant with respect to $G$ if:

$$f(T(g)x) = T'(g)f(x), \forall g \in G, x \in X \tag{4}$$

Here, $T(g)$ and $T'(g)$ describe how the group elements $g$ act on the input and output spaces. A special case of equivariance is invariance, where $T'$ reduces to the identity transformation.

**From Rotational Symmetry of Systems to Cyclic Group Equivariance**

Many CT and ECT systems exhibit rotational symmetry about their central axis, a property arising from the circular arrangement or periodic rotation of detectors. To simplify the analysis, we assume the system is effectively time-invariant during the entire imaging process, under which the discussions for both static arrangements and rotational motion are equivalent. For clarity, we consider a system with $N$ identical detector arrays uniformly distributed in a clockwise arrangement around the central axis.

Let $\boldsymbol{h_0}(\mathbf{r})$ denote the sensitivity function of the detector array at the initial position (the $0\,th$ detector), where $\mathbf{r}$ represents the three-dimensional position vector from the viewpoint of the $0\,th$ detector. We define a rotation operator $R_n$ representing a counterclockwise rotation by an angle of $\frac{2n\pi}{N}$ about the central axis, where $n \in \mathbb{Z}_N$. These rotations form a cyclic group $C_N^S$ defined on the object space $S = \mathbb{R}^3$. The $n\,th$ detector is identical to the $0\,th$ detector, with its position derived by applying $R_n$ to the $0\,th$ detector. Therefore, the sensitivity function of the $n\,th$ detector, $\boldsymbol{h_n}(\boldsymbol{r})$ can be related to that of the $0\,th$ detector as follows:

$$\boldsymbol{h_n}(\mathbf{r}) = \boldsymbol{h_0}(R_n^{-1}\mathbf{r}) = \boldsymbol{h_0}(R_{N-n}\mathbf{r}),\ n \in \mathbb{Z}_N. \tag{5}$$

Assuming the object function in the object space $S$ is $y(\mathbf{r})$, which describes the signal distribution of the scanned object and $y(\mathbf{r}) \in \mathbb{R}$. The measured data of the entire system is defined in the projection space $P = \mathbb{Z}_N \times \mathbb{R}^d$, where $\mathbb{Z}_N$ represents the discrete angular positions and $\mathbb{R}^d$ corresponds to the detector readings at each position. The data is represented as $\boldsymbol{g} = [\boldsymbol{g_0}, \boldsymbol{g_1}, \ldots, \boldsymbol{g_{N-1}}]$, with each $\boldsymbol{g_n} \in \mathbb{R}^d$ denoting the projection data collected by the $n$ th detector array. Assume $\mathcal{T}_0$ is a right cyclic rotation operator in $P$, defined as

$$\mathcal{T}_0 \boldsymbol{g} = [\boldsymbol{g_{N-1}}, \boldsymbol{g_0}, \ldots, \boldsymbol{g_{N-2}}]. \tag{6}$$

This operator performs a right cyclic rotation along the discrete angle axis $\mathbb{Z}_N$ of the projection space $P$. Using $\mathcal{T}_0$ as generator, we define a cyclic group $C_N^P = \langle \mathcal{T}_0 \rangle$ where $\mathcal{T}_n = \mathcal{T}_0^n$ represents the right cyclic rotation applied $n$ times. To encompass various imaging modalities and geometric configurations, the forward projection process is described as:

$$g_n = \int_S h_n(\mathbf{r})y(\mathbf{r})d^3\mathbf{r}, \; n \in \mathbb{Z}_N. \tag{7}$$

Let $\mathbf{r}'$ be the three-dimensional position vector from the viewpoint of the $m\,th$ detector ($m \in \mathbb{Z}_N$). The coordinate transformation is given by $\mathbf{r} = R_m\mathbf{r}'$. With $\mathbf{r}'$ denotes the transformed coordinates, the sensitivity function of the $n\,th$ detector is $h_n(R_m\mathbf{r}')$. When the object is rotated by $R_m$, its corresponding object function $y(\mathbf{r})$ is transformed into $y(\mathbf{r}')$. The corresponding measured data for the $n\,th$ detector is denoted as $g'_n$. From (7), we have

$$g'_n = \int_S h_n(R_m\mathbf{r}')y(\mathbf{r}')d^3\mathbf{r}', \; m, n \in \mathbb{Z}_N. \tag{8}$$

Based on the detector's relationship from (5), the sensitivity function satisfies

$$h_n(R_m\mathbf{r}') = h_q(\mathbf{r}), \tag{9}$$

where $q = (n - m) \bmod N$, and $q \in \mathbb{Z}_N$. Consequently, the measured data satisfies $g'_n = g_q$. This implies that the overall measured data $g$ before and after the rotation are related as follows:

$$g' = \mathcal{T}_m g, \; m \in \mathbb{Z}_N. \tag{10}$$

Initially, $\mathcal{T}_m$ was just defined as a purely mathematical concept. Equitation (10) highlights the practical significance of it: *after acting on $g$, the transformed $g'$ still represents the same physical object, albeit rotated*. It imposes a uniformly distributed circular structure on the projection space $P$, reflecting the inherent rotational properties of the system.

The reconstruction process can be regarded as the inverse of the imaging process described above. From the perspective of the entire imaging pipeline, symmetry in the projection space $P$ is inherently linked to symmetry in the object space $S$. Let $\mathcal{F}$ denote the reconstruction mapping from $P$ to $S$. We obtain the following properties of the mapping

$$\mathcal{F}[\mathcal{T}_m g] = R_m[y], \; \forall\, m \in \mathbb{Z}_N, g \in P. \tag{11}$$

Equation (11) demonstrates that $\mathcal{F}$ exhibits equivariance under the cyclic group $C_N$, which is defined by the rotational symmetry of the systems. Specifically, the cyclic group has two corresponding representations in their respective spaces: cyclic rotations in the projection domain and discrete rotation in the object domain. This property is referred to as *cyclic group equivariance* (CGE).

**Cyclic Rotation Equivariant Convolution**

To preserve the inherent circular structure of the projection domain $P$, we propose a novel design for cyclic rotation equivariant convolution (CREC). This design builds upon the group-convolution framework, adapted to the cyclic group structure that arises from hardware-induced rotational symmetry.

To lay the foundation, we first introduce the concept of group convolution, a generalization of standard convolution that has been widely applied in tasks requiring symmetry-preserving operations [14]. Formally, let $G$ be a group, and let $f: G \to V$ represent an input feature map ($V$ being the feature space). A convolutional kernel $\kappa$ is defined as a mapping from the group to homomorphisms between $V$ and the output feature space $W$. Formally, group convolution is defined as:

$$[\kappa \star f](g) = \int_G \kappa(g^{-1}h)f(h)dh \tag{12}$$

where $dh$ is a left-invariant Haar measure on G, and the integral is taken over the group $G$. It ensures that when the input transforms under $G$, the output transforms correspondingly, preserving the symmetry defined by the group.

A familiar example of this is the standard convolution operation in CNNs, where the group $\mathbb{Z}^2$ represents integer translations. For discrete inputs like 2D images ($f: \mathbb{Z}^2 \to \mathbb{R}$), the integral reduces to a summation:

$$[\kappa \star f](x) = \sum_{u \in \mathbb{Z}^2} \kappa(x - u)f(u), \tag{13}$$

ensuring translational equivariance: spatial shifts in the input result in corresponding shifts in the output feature maps.

Extending the group convolution concept to projection data, we employ the cyclic rotation group $C_N^P$. The input projection data is represented as $\boldsymbol{g}(n): \mathbb{Z}_N \to \mathbb{R}^d$, where $\boldsymbol{g}(n)$ corresponds to the projection at the $n$-th angular position, the cyclic rotation convolution is defined as:

$$[\kappa \star \boldsymbol{g}](m) = \sum_{n \in \mathbb{Z}_N} \kappa(m - n)\boldsymbol{g}(n) \tag{14}$$

This formulation ensures that when the input undergoes a cyclic rotation $\mathcal{T}_n$, the output transformed accordingly, maintaining equivariance under $C_N^P$. By applying $t$ convolutional kernels $\kappa$, we obtain the output features $\hat{\boldsymbol{g}}(n): \mathbb{Z}_N \to \mathbb{R}^t$, which inherit a circular data structure similar to $\boldsymbol{g}(n)$.

The core of the practical implementation of group convolution lies in weight sharing across group transformations. For cyclic rotation group, this corresponds to a circular shift in the arrangement of projections along the angular axis, which can be achieved by adapting the standard convolution operation. The operation is confined to $C_N^P$ by restricting shifts to the angular axis with a stride of 1, ensuring all group elements are covered. To preserve the circular structure of the angular axis, the kernel needs to wrap around at the boundaries of projections, maintaining periodicity. This is critical when the kernel width along the angular dimension exceeds 2, which can be achieved by padding the edge data appropriately. Fig. 2 illustrates one simple implementation, where the kernel width is set to 1, eliminating the need for wrapping around.

As the experiments in this study focus on domain transform reconstruction, we integrate three CREC layers into the AUTOMAP backbone. AUTOMAP [3], a representative model for domain transformation reconstruction, is composed of two fully connected layers of considerable size, followed by an autoencoder constructed with convolutional layers. As shown in Fig. 3, placing the CREC layers at the input stage ensures that the symmetric structure of the projection data is preserved from the outset, providing more reliable feature representation. To maintain sufficient representational capacity, the input and output dimensions of the CREC layers are designed to remain consistent.

**Model Design and Cyclic Group Equivariance Regularization**

To ensure cyclic equivariance throughout the entire reconstruction process, we introduce Cyclic Group Equivariance Regularization (CGER). Whereas CREC imposes symmetry in the projection domain by design, CGER reinforces consistent transformations across the entire network from an optimization perspective. CGER achieves this by penalizing deviations from equivariance during the optimization of reconstruction network, as quantified by the following regularized objective function:

$$\arg\min_{\theta} \frac{1}{N-1} \sum_{i=1}^{N-1} \|\mathcal{F}(T_i \boldsymbol{g}, \boldsymbol{\theta}) - R_i \mathcal{F}(\boldsymbol{g}, \boldsymbol{\theta})\|^2, \quad (14)$$

Where $\boldsymbol{\theta}$ represents the parameters of the reconstruction network $\mathcal{F}$, and $\mathcal{F}(\boldsymbol{g})$ is the reconstructed image from the projection data $\boldsymbol{g}$. As discussed earlier, the equivariance is defined with respect to the cyclic group $C_N$, which represents N-fold rotational symmetry of the system.

However, directly regularizing across the entire group $C_N$ can be computationally expensive when $N$ is large. To address this, a practical approach is to use a subgroup $C_M \subseteq C_N$, where M is divisor of $N$. While $C_M$ contains fewer elements than $C_N$, it can still capture a significant degree of the original symmetry, providing a balance between computational efficiency and maintaining the equivariance constraints.

# Experiments and Results

In this section, we evaluate the proposed methods on diverse brain phantom datasets to investigate their ability to generalize across different data distributions, and to characterize how the proposed CREC and CGER components contribute to reconstruction quality and stability.

## Datasets

A reliable reconstruction model is expected to perform effectively against deviations in data distributions while accurately capturing the latent relationships between the projection domain and the image domain. To address this, we designed experiments using two types of digital phantoms based on 20 anatomical brain models from BrainWeb [27]. Each model contains discrete phantoms, characterized by tissue boundaries and uniform pixel distributions, and fuzzy phantoms, which exhibit blurred boundaries and more realistic pixel distributions corresponding to specific brain tissues.

Among the 20 models, 17 discrete phantoms were used for training, 1 discrete phantom was reserved for validation, and 2 discrete phantoms were used as the standard test set. Additionally, the fuzzy phantoms derived from the two test models, specifically cerebrospinal fluid (CSF), white matter (WM), and gray matter (GM), were employed as external test datasets. Testing on the fuzzy phantoms provided a critical evaluation of the models' generalization, as these phantoms introduced significant deviations in data distributions compared to the discrete phantoms used during training.

During preprocessing, the 3D phantoms were converted into 2D slices, and 290 slices were retained from each volume. Each slice was downsampled and padded into a square size of $190 \times 190$. Given the limited training dataset size ($17 \times 290$), data augmentation was applied, including rotations in 90° increments (90°, 180°, and 270°) and flips. These operations expanded the training set to $17 \times 290 \times 8$ samples. To simulate forward projections, the radon transform was applied to the preprocessed phantoms, generating projection matrices of size $195 \times 128$, corresponding to 128 projection angles evenly distributed around a full circle.

**Implementation Details**

The loss function in our experiments consisted of two components: the main loss $\mathcal{L}_{\text{main}}$, defined as the mean squared error (MSE) between the reconstructed output and the ground truth, and the CGER constraint $\mathcal{L}_{\text{CGER}}$, as described in (14). Rotational transformations in the image domain were performed using bicubic interpolation. These two components were combined as:

$$\mathcal{L} = \mathcal{L}_{\text{main}} + \lambda \cdot \mathcal{L}_{\text{CGER}} . \tag{15}$$

To balance between the reconstruction mapping and the CGER constraint, the weight $\lambda$ was dynamically adjusted at each batch iteration. Specifically, $\lambda$ was computed as:

$$\lambda = \frac{\mathcal{L}_{\text{CGER}}.detach}{\mathcal{L}_{\text{main}}.detach} \tag{16}$$

Where $.detach$ ensures that the gradient flow is not influenced by the ratio computation itself, thus maintaining stability. This design functions as a feedback mechanism: if the ratio $\frac{\mathcal{L}_{\text{CGER}}}{\mathcal{L}_{\text{main}}}$ is large, $\lambda$ increases to strengthen the equivariance constraint; otherwise, $\lambda$ decreases to prioritize reconstruction.

To manage computational resources, we chose $M = 32$, as the subgroup of $C_{128}$. Other hyperparameters included a batch size of 44, a learning rate of 0.00001, and the ADAM optimizer. All experiments were conducted on a workstation equipped with an NVIDIA A800 GPU. We evaluated four model configurations: the original AUTOMAP and the modified AUTOMAP (with CREC layers), each trained with and without the CGER constraint.

**Training and Convergence Analysis**

Fig. 4 shows the training curves for the original AUTOMAP model and the modified AUTOMAP (i.e., with CREC layers). Compared to the original model, the modified AUTOMAP demonstrates more stable convergence with fewer oscillations, irrespective of whether the CGER constraint is applied. Meanwhile, adding CGER reduces the gap between training and validation curves in both models, suggesting improved generalization and reduced overfitting. Importantly, the CGER constraint does not significantly alter overall convergence speed, as both models require a similar number of epochs to achieve their respective plateaus.

**Discrete Anatomical Model Results**

To evaluate reconstruction quality, we use three metrics: Structural Similarity Index Measure (SSIM), Peak Signal-to-Noise Ratio (PSNR), and relative Root Mean Square Error (rRMSE). TABLE I presents the quantitative results, and the Filtered Back Projection (FBP) algorithm is included as a conventional reconstruction baseline for comparison. Fig. 5 illustrates representative reconstructions, showing that all deep learning-based methods produce smoother images and effectively suppress streak artifacts compared to FBP, as reflected by higher SSIM values.

First, we examine the impact of CREC layers. The original AUTOMAP tends to preserve fewer fine details, as indicated by its relatively lower PSNR and higher rRMSE than FBP, likely due to the limited diversity of the training set. In contrast, the modified AUTOMAP with CREC exhibits superior performance, consistently outperforming FBP across all metrics and striking a better balance between artifact suppression and detail preservation. This highlights the effectiveness of CREC in extracting reliable projection domain features that are consistent with the underlying rotational symmetry.

Next, we consider the effect of the CGER constraint on the two models. For the original AUTOMAP, CGER yields clear benefits: SSIM increases by 0.0141 and PSNR by 0.38 dB, corresponding to visually noticeable improvements. However, for the modified AUTOMAP, the CGER constraint shows minimal additional gains (an SSIM increase of 0.0045 and a PSNR drop of 0.05 dB). When tested on phantoms similar to those seen during training, the CGER helps the original AUTOMAP but has limited impact on the already more efficient modified version.

**Fuzzy Phantom Model Results**

We also assessed model performance on fuzzy phantom datasets derived from the same two test brain models. As reported in TABLE I, all methods exhibit a performance drop, as these fuzzy phantoms introduce significant distribution shifts relative to the discrete phantoms used for training. Notably, FBP also degrades substantially, likely due to the higher frequency components in the fuzzy structures.

The CGER constraint substantially improves performance on these fuzzy datasets, particularly for the original AUTOMAP. As shown in Fig. 6–8, constrained models deliver more detailed and stable reconstructions, mitigating severe artifacts and structural distortions. SSIM increases by 0.2012 (GM), 0.2150 (WM), and 0.3114 (CSF) under CGER for the original AUTOMAP, indicating a significant leap in visual quality. For the modified AUTOMAP, the SSIM improvements are also notable—0.1553 (GM), 0.1680 (WM), and 0.1862 (CSF)—though slightly smaller than in the original model, reflecting an already higher baseline stability.

Regarding CREC layers, even without CGER, the modified AUTOMAP shows better robustness compared to the original AUTOMAP, as evidenced by reduced blurring and lighter artifacts in Fig. 6–8 and by the quantitative results in Table I. Specifically, SSIM improves by 0.0431 (GM), 0.0508 (WM), and 0.1042 (CSF), while PSNR gains are 2.25 dB, 3.25 dB, and 3.54 dB, respectively. These differences in performance highlight the role of CREC in inherently introducing symmetry-awareness; adding CGER on top of CREC yields further, albeit smaller, improvements for the modified model.

**Evaluation of the Symmetry Consistency**

Finally, to assess rotational consistency under the full cyclic group $C_{128}$, we extended the CGER constraint as an evaluation metric. Specifically, we computed the rRMSE between $\mathcal{F}(T_j \boldsymbol{g}, \boldsymbol{\theta})$ and $R_j \mathcal{F}(\boldsymbol{g}, \boldsymbol{\theta})$ for all $j = 1 \dots 127$ (excluding the identity at 0°) and aggregated these results across all four test sets. The rRMSE values for each angular transformation were plotted in a polar coordinate diagram, providing a comprehensive visualization of each model's consistency with the $C_{128}$ symmetry.

Fig. 9 shows that constrained models achieve lower and more uniform error distributions across all angles, demonstrating enhanced rotational symmetry compared to unconstrained versions. Notably, the modified AUTOMAP model with CGER achieves the lowest and most consistent angular errors distributions across all datasets, underscoring the impact of symmetry-aware design proposed in this study. Minor anomalies at specific angles (90°, 180°, 270°) are likely attributable to interpolation errors introduced by the rotation operations.

Overall, these findings confirm that enforcing cyclic group equivariance—both through architecture (CREC) and training regularization (CGER)—helps maintain rotational consistency and improves the robustness of the reconstruction process.

**Discussion and Conclusion**

In this study, we revisited the reconstruction process from the perspective of hardware-induced rotational symmetry and introduced the concept of cyclic group equivariance. For deep learning-based reconstruction, this concept offers a fundamental principle to consider for model design and optimization, independent of specific projection geometries. While using rotational symmetry in deep learning-based reconstruction is novel, its application in conventional handcrafted (i.e., non-learning-based) reconstruction methods is well established. For example, iterative reconstruction algorithms in CT, PET, and OAT often exploit rotational symmetry to simplify system matrix modeling, substantially reducing computational costs [28]–[32]. The proposed approach extends this idea to data-driven reconstruction methods by incorporating equivariance, thereby enhancing model interpretability and reliability.

Our experiments primarily focus on domain-transform reconstruction, proposing two complementary designs to achieve cyclic group equivariance: (1) a CREC layer applied to the projection domain for reliable feature extraction (i.e., network design perspective), and (2) a CGER constraint applied across the entire network (i.e., network optimization perspective). One notable challenge in deep learning-based reconstruction is the reliance on limited datasets, which often leads to instability when encountering data distributions that differ from training conditions. We addressed this concern by training on relatively homogeneous discrete brain phantoms and then testing on fuzzy phantoms with more complex intensity distributions. The results underscore the benefits of both CREC and CGER in enhancing model stability. Specifically, on standard discrete phantom test sets, CREC layers markedly improve reconstruction performance, while CGER exerts a more pronounced effect on external fuzzy phantom test sets, substantially boosting stability and generalizability. Moreover, the modified model incorporating CREC consistently outperforms the original AUTOMAP, highlighting its ability to learn reliable representations from limited training data. Although the current framework does not realize strict equivariance through the entire network—given the complexity of implementing cross-domain symmetry—our symmetry consistency evaluations reveal that both CREC and CGER introduce substantial equivariant behavior, reflected by more uniform and compact error distributions.

It is important to note that cyclic group equivariance is not confined to domain-transform reconstruction alone. As a manifestation of the system's inherent rotational symmetry, it naturally emerges in the projection domain as well. Equation (10) demonstrates the projection domain's equivariance with respect to $C_N^P$, emphasizing the circular structure of its angular dimension $\mathbb{Z}_N$. This structure arises fundamentally from hardware-induced symmetry, independent of specific data characteristics, and it motivates the CREC layer design. Beyond pure domain-transform approaches, any projection-domain deep learning task or hybrid reconstruction method that processes projection data stands to benefit from the CREC methodology proposed here.

All discussions in this work assume strict rotational symmetry. However, real-world systems may exhibit non-idealities—such as hardware sensitivity variations or scatter artifacts—that break perfect symmetry. While hardware corrections (e.g., sensitivity calibration, scatter correction) can address some issues, in more severe cases the symmetry might be irrevocably compromised. Still, a more common scenario involves partial or relaxed symmetry. Our proposed methods, both in network design and in optimization, constitute a flexible framework for such scenarios. For instance, adjusting the weight of CGER can relax the equivariance requirements to accommodate moderate symmetry-breaking factors. Alternatively, combining equivariant layers with other model components can selectively reintroduce symmetry into parts of the reconstruction pipeline, offering a practical balance between theoretical rigor and real-world feasibility.

The primary goal of this work is to introduce a unified framework for symmetry-aware design and highlight its potential advantages. Consequently, the models and experiments presented here are not tailored to one specific clinical setting. While we focus on rotational symmetry in the angular dimension $\mathbb{Z}_N$ of the projection space $\mathbb{Z}_N \times \mathbb{R}^d$, other forms of prior knowledge—such as those pertaining to the $\mathbb{R}^d$ subspace —could further strengthen reconstruction stability. Clearly, priors related to the subspace $\mathbb{R}^d$ do not conflict with symmetry principles. Rather, integrating them with symmetry-aware designs could ultimately yield more reliable and interpretable models.

Lastly, the methods proposed here for incorporating cyclic group equivariance into deep learning warrant further exploration. While the CGER framework can be adapted to various neural architectures, it does increase computational complexity. Similarly, the CREC design effectively enforces strict equivariance in the projection domain but may be restrictive if used as the sole mechanism for symmetry. Future research could consider alternative weight-sharing strategies and expanded CREC-like designs that address the entire cyclic group $C_N^P$, leading to more flexible and efficient symmetry-aware networks. We believe that our introduction of cyclic group equivariance principle can help lay a solid foundation for the future development of stable and interpretable deep learning-based reconstruction methods in a range of imaging systems featuring rotational symmetry.

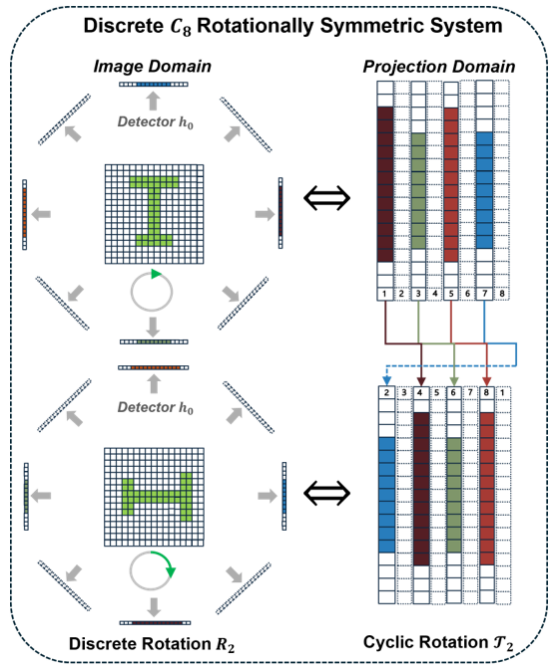

*Figure 1. Discrete $C_8$ Rotationally Symmetric System. Illustration of how rotational symmetry is manifested in both the image domain (left) and the projection domain (right) for an $N = 8$ tomographic configuration. A discrete rotation $R_2$ in the image domain corresponds to a cyclic rotation $\mathcal{T}_2$ of the projections, both governed by the same cyclic group $C_8$. This highlights the cross-domain nature of rotational symmetry, forming the basis of cyclic group equivariance.*

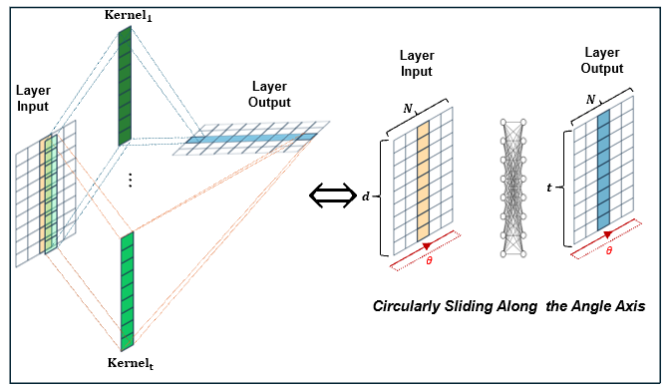

*Figure 2. Conceptual Diagram of the proposed CREC Layer operating on Projection Data. Multiple kernels convolve along the angular axis, wrapping around the boundary to maintain circularity. By enforcing equivariance with respect to cyclic rotation group $C_N^P$ in the projection domain, CREC captures the intrinsic rotational symmetry imposed by the hardware design.*

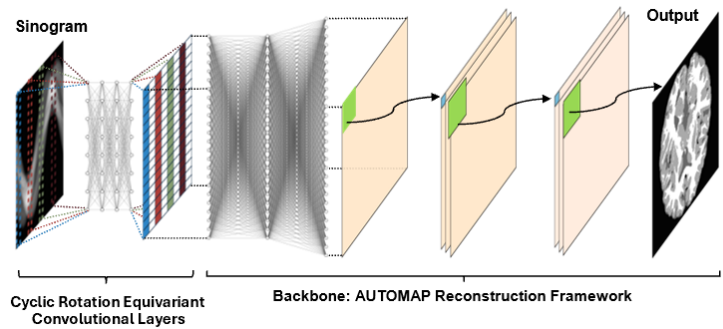

*Figure 3. Overall Reconstruction Framework (modified AUTOMAP with three CREC layers). CREC layers handle cyclic rotation features along the angle dimension, preserving symmetry-aware representations.*

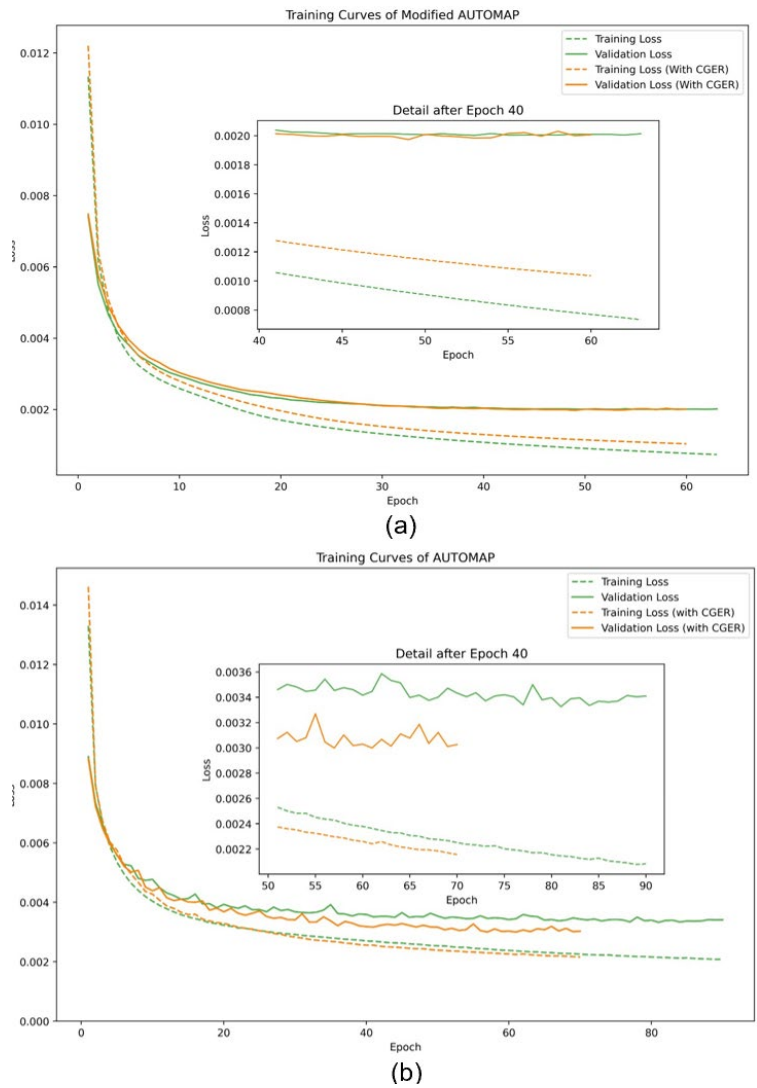

*Figure 4. Comparison of training loss (solid lines) and validation loss (dashed lines) for: modified AUTOMAP with CREC(a), and original AUTOMAP model (b). In both figures, the training loss and validation loss are plotted against epochs. Insets provide a close-up view after epoch 40, illustrating how CGER stabilizes convergence and narrows the training–validation gap.*

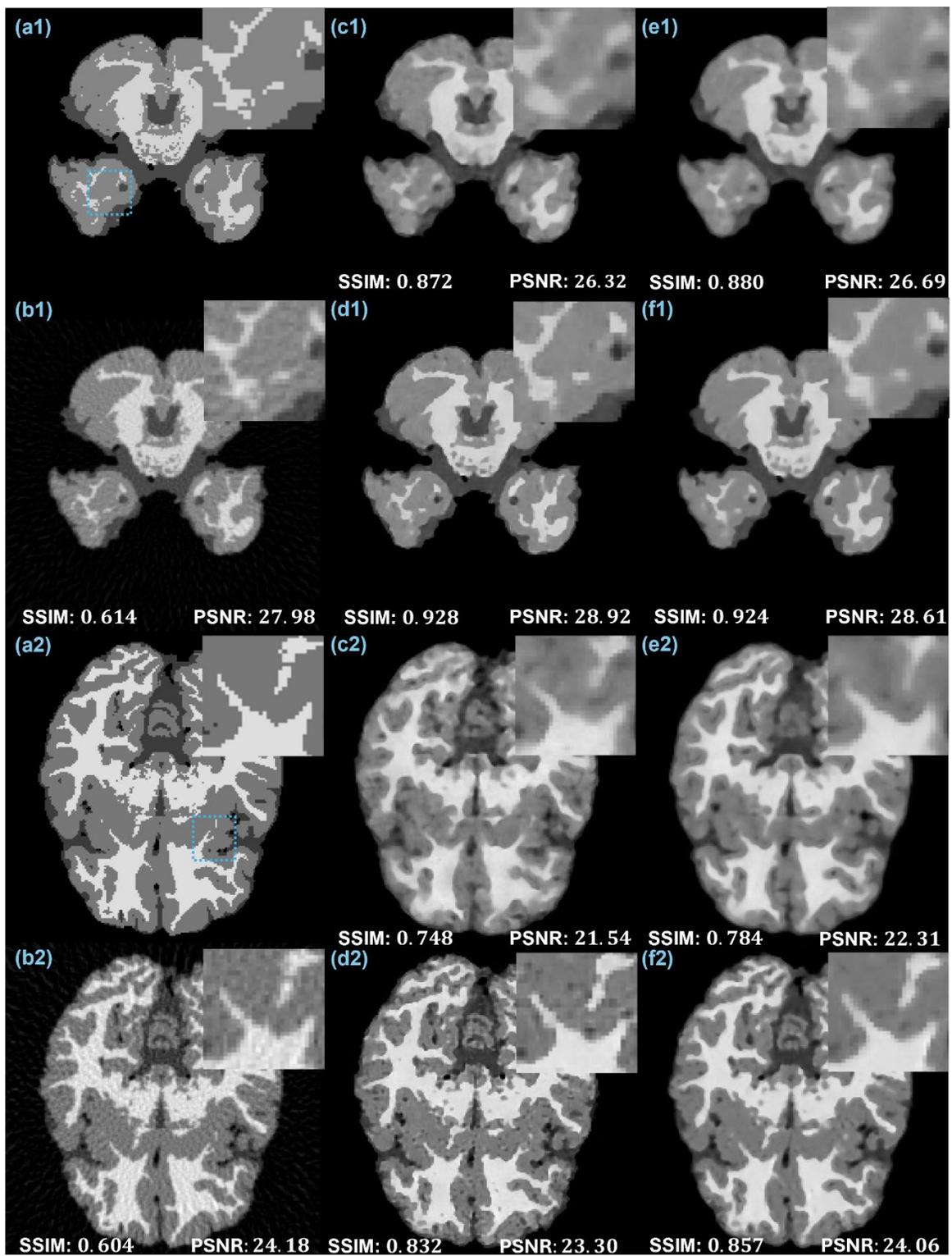

*Figure 5. Reconstruction results on discrete phantom dataset: (a) Ground Truth; (b) FBP; (c) Original AUTOMAP; (d) Modified AUTOMAP; (e) AUTOMAP with CGER constraint; and (f) Modified AUTOMAP with CGER constraint. SSIM and PSNR values are annotated for each method. 1 and 2 represent the different slices.*

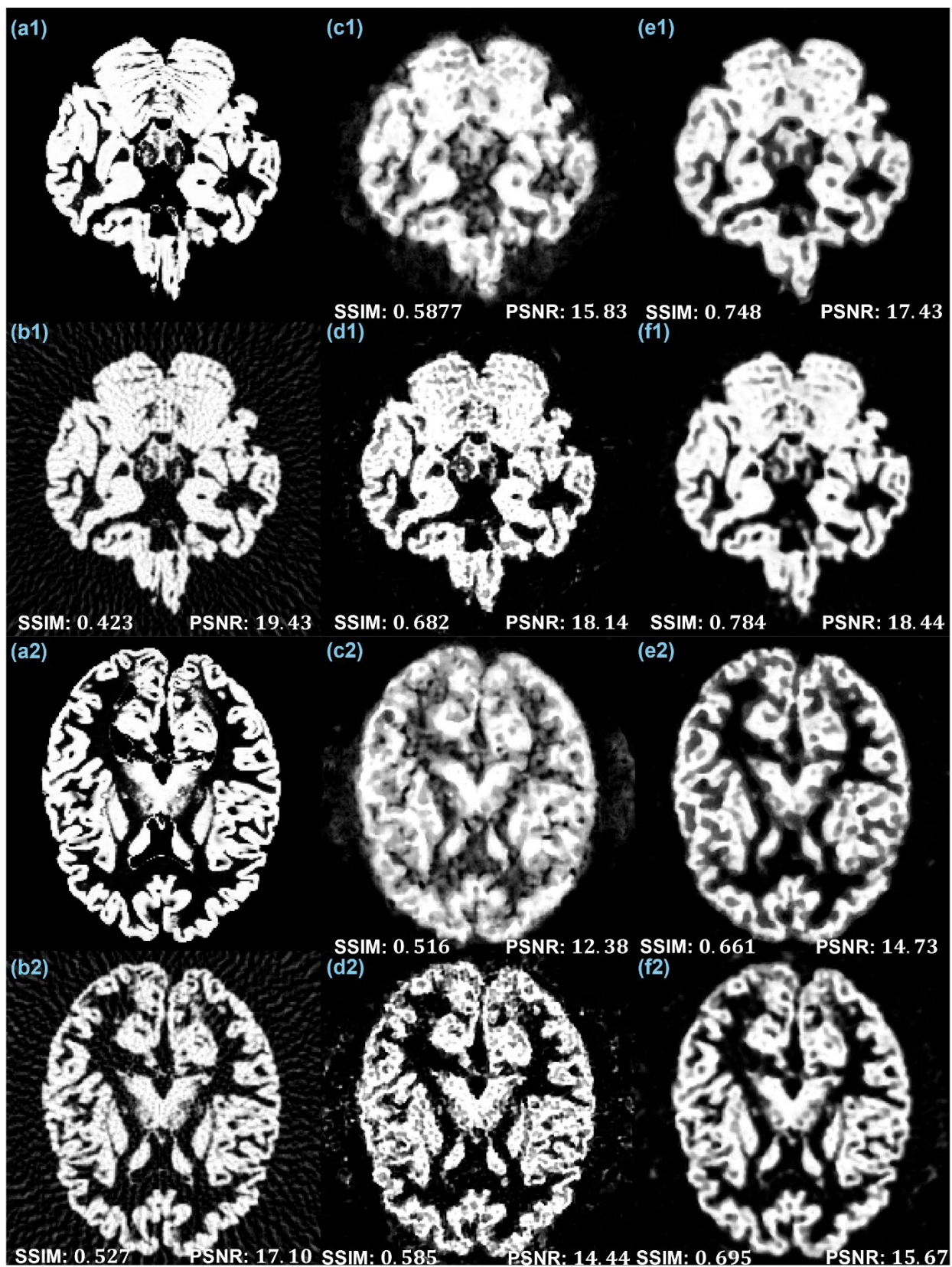

*Figure 6. Reconstruction Results on Grey Matter (GM) Fuzzy Phantoms. (a) Ground Truth; (b) FBP; (c) Original AUTOMAP; (d) Modified AUTOMAP; (e) AUTOMAP with CGER constraint; and (f) Modified AUTOMAP with CGER constraint. SSIM and PSNR values are annotated for each method. 1 and 2 represent the different slices.*

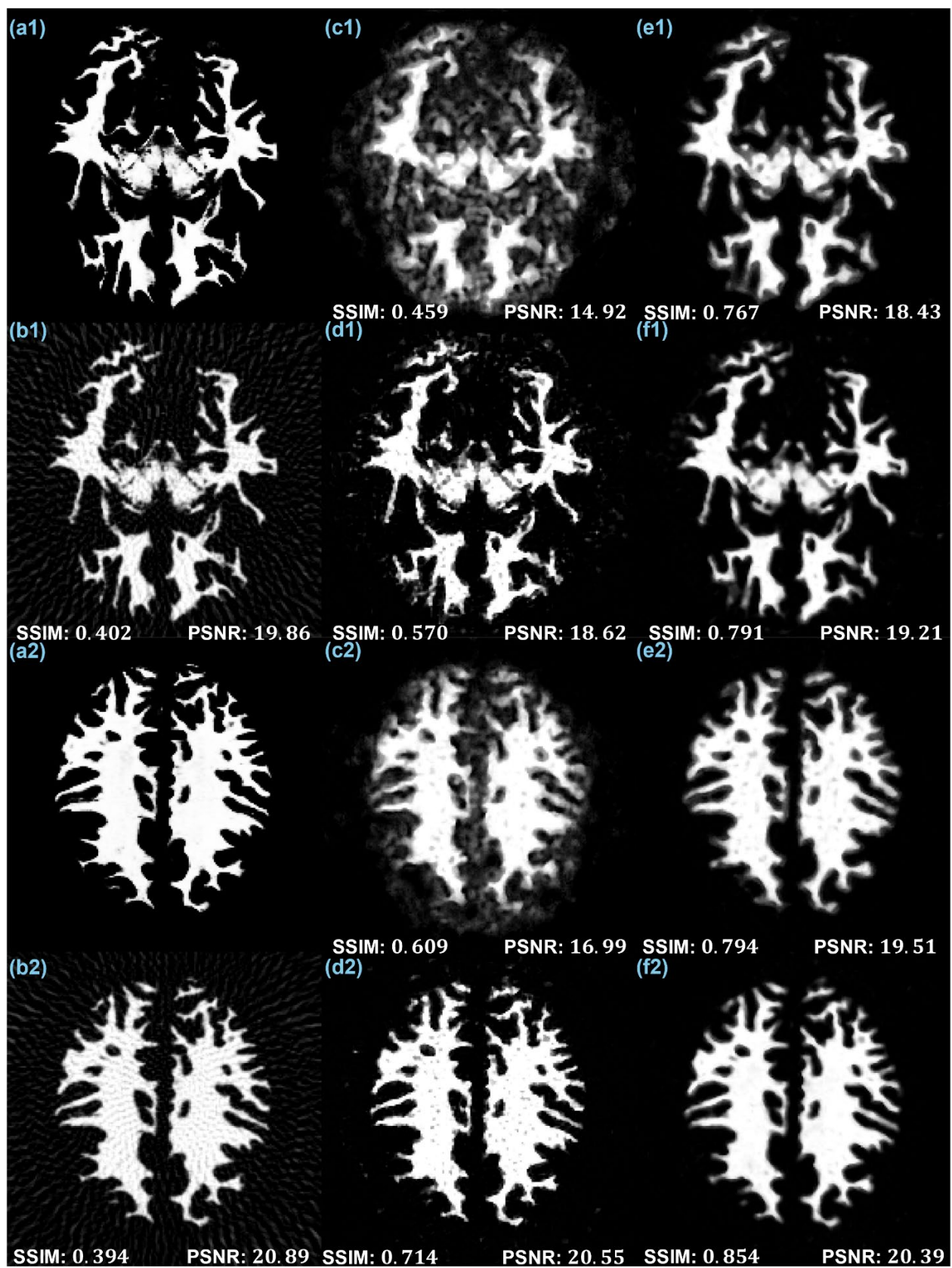

*Figure 7. Reconstruction Results on White Matter (WM) Fuzzy Phantoms. (a) Ground Truth; (b) FBP; (c) Original AUTOMAP; (d) Modified AUTOMAP; (e) AUTOMAP with CGER constraint; and (f) Modified AUTOMAP with CGER constraint. SSIM and PSNR values are annotated for each method. 1 and 2 represent the different slices.*

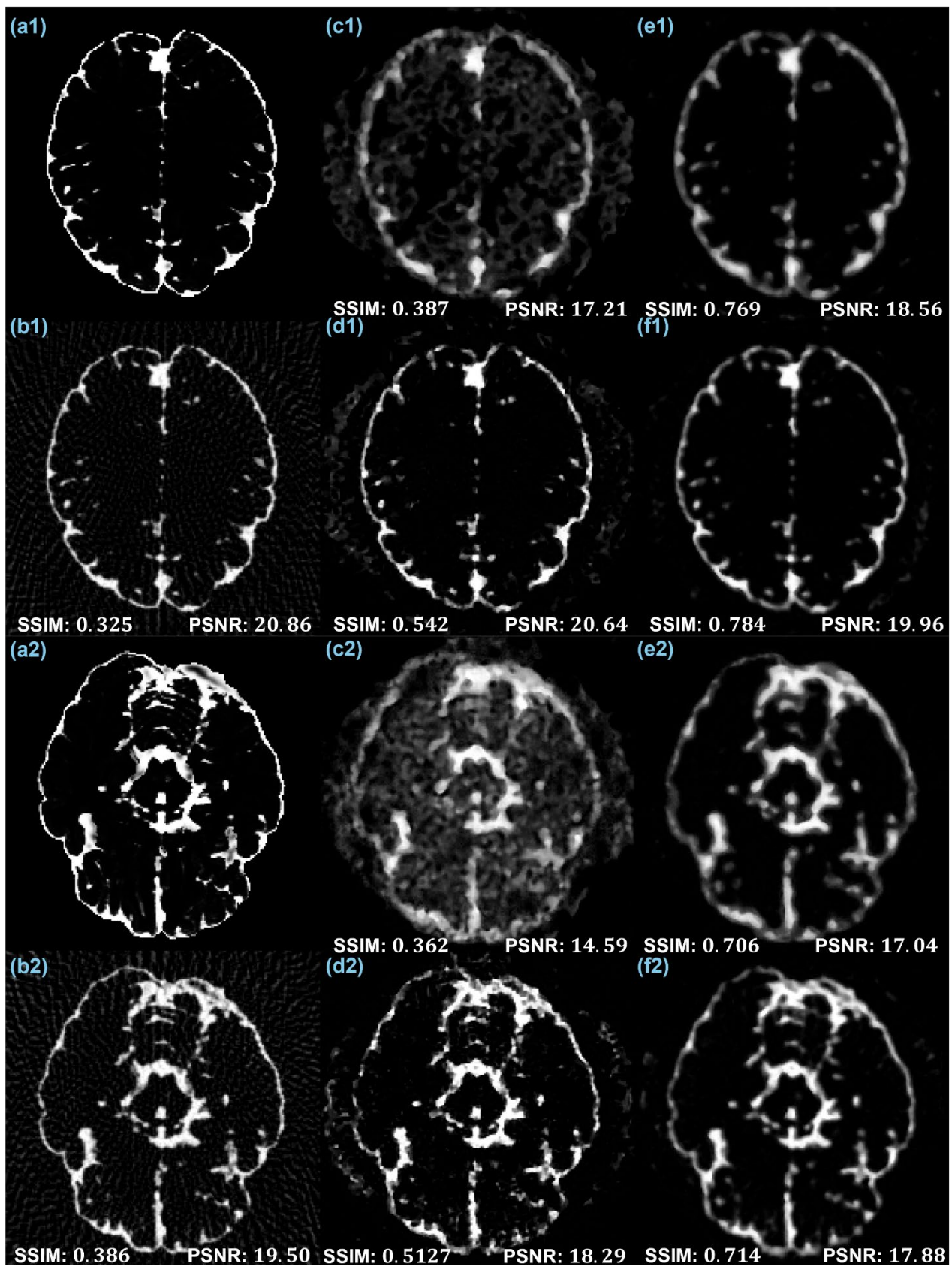

*Figure 8. Reconstruction Results on cerebrospinal fluid (CSF) Fuzzy Phantoms. (a) Ground Truth; (b) FBP; (c) Original AUTOMAP; (d) Modified AUTOMAP; (e) AUTOMAP with CGER constraint; and (f) Modified AUTOMAP with CGER constraint. SSIM and PSNR values are annotated for each method. 1 and 2 represent the different slices.*

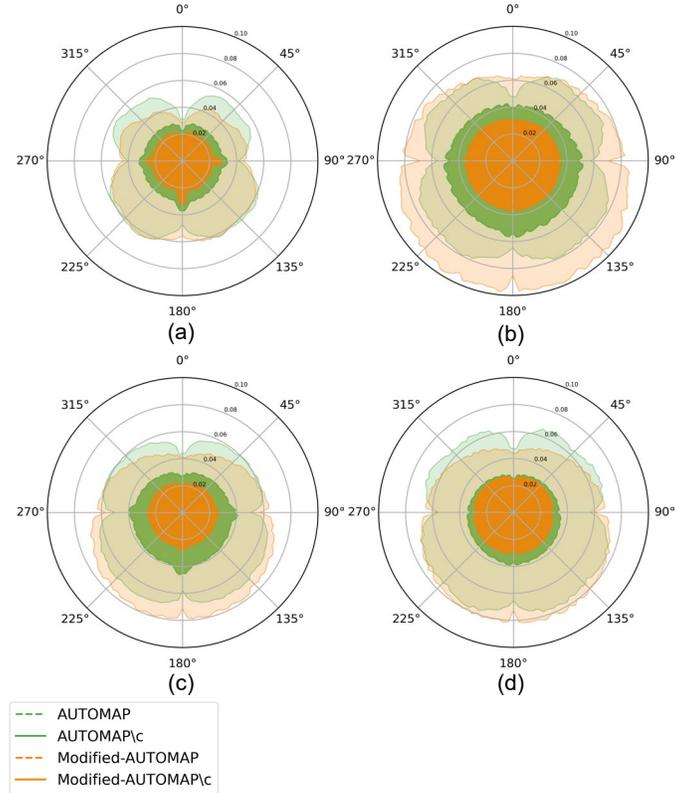

*Figure 9. Polar Plots for Symmetry Consistency. (a) Results for the discrete phantom. (b), (c), and (d) show results for the fuzzy phantom corresponding to grey matter (GM), white matter (WM), and cerebrospinal fluid (CSF), respectively. Polar coordinate plots illustrating the relative RMS error (rRMSE) across 127 angular transformations (excluding the identity) for each reconstruction model. Smaller radii and more uniform distributions indicate stronger symmetry consistency.*

TABLE I. Quantitative Comparison of Different Reconstruction Methods.

| | SSIM | PSNR | rRMSE |
|---|---|---|---|
| **Discrete Model** | | | |
| FBP | 0.6492 ± 0.0618 | 27.84 ± 2.63 | 0.1255 ± 0.0220 |
| AUTOMAP | 0.8506 ± 0.0759 | 25.92 ± 3.22 | 0.1555 ± 0.0197 |
| AUTOMAP/c | 0.8647 ± 0.0667 | 26.30 ± 2.98 | 0.1491 ± 0.0210 |
| Mod-AUTOMAP | 0.9144 ± 0.0473 | 28.29 ± 3.32 | 0.1186 ± 0.0172 |
| Mod-AUTOMAP/c | 0.9189 ± 0.0398 | 28.34 ± 3.02 | 0.1182 ± 0.0186 |
| **Fuzzy Model: Grey Matter** | | | |
| FBP | 0.4257 ± 0.0743 | 19.83 ± 3.08 | 0.2720 ± 0.0492 |
| AUTOMAP | 0.5664 ± 0.1139 | 16.45 ± 3.96 | 0.4022 ± 0.0731 |
| AUTOMAP/c | 0.7676 ± 0.1160 | 18.01 ± 3.46 | 0.3356 ± 0.0575 |
| Mod-AUTOMAP | 0.6095 ± 0.0937 | 18.70 ± 4.36 | 0.3123 ± 0.0642 |
| Mod-AUTOMAP/c | 0.7648 ± 0.1075 | 18.96 ± 3.83 | 0.3012 ± 0.0516 |
| **Fuzzy Model: White Matter** | | | |
| FBP | 0.4276 ± 0.1066 | 22.98 ± 3.77 | 0.2958 ± 0.0763 |
| AUTOMAP | 0.6321 ± 0.1734 | 19.98 ± 5.04 | 0.4113 ± 0.0687 |
| AUTOMAP/c | 0.8471 ± 0.0876 | 21.62 ± 3.78 | 0.3462 ± 0.0921 |
| Mod-AUTOMAP | 0.6829 ± 0.1165 | 23.23 ± 4.43 | 0.2853 ± 0.0634 |
| Mod-AUTOMAP/c | 0.8509 ± 0.0792 | 22.79 ± 4.38 | 0.2887 ± 0.0621 |

| Fuzzy Model: Cerebrospinal Fluid | | | |
|---|---|---|---|
| FBP | 0.3543 ± 0.0308 | 20.96 ± 1.71 | 0.3793 ± 0.0497 |
| AUTOMAP | 0.4756 ± 0.1496 | 17.28 ± 2.85 | 0.5862 ± 0.1179 |
| AUTOMAP/c | 0.7870 ± 0.0928 | 18.86 ± 1.97 | 0.4846 ± 0.0718 |
| Mod-AUTOMAP | 0.5798 ± 0.0771 | 20.82 ± 2.62 | 0.3886 ± 0.0706 |
| Mod-AUTOMAP/c | 0.7660 ± 0.0871 | 19.70 ± 2.37 | 0.4395 ± 0.0632 |